\documentclass[pra,twocolumn,floatfix]{revtex4}
\usepackage{graphicx}
\usepackage{color}
\usepackage{amsmath}
\begin{document}

\title{Phase diagram of rotating Bose-Einstein condensates trapped in power-law and hard-wall 
potentials}
\author{G. M. Kavoulakis$^{1,2}$}
\affiliation{$^1$Department of Mechanical Engineering, Hellenic Mediterranean University, 
P.O. Box 1939, GR-71004, Heraklion, Greece
\\
$^2$HMU Research Center, Institute of Emerging Technologies, GR-71004, Heraklion, Greece}
\date{\today}

\begin{abstract}

We investigate the rotational phase diagram of a quasi-two-dimensional, weakly-interacting 
Bose-Einstein condensate confined in power-law and in hard-wall trapping potentials. For weak 
interactions, the system undergoes discontinuous transitions between multiply-quantized 
vortex states as the rotation frequency of the trap increases. In contrast, stronger 
interactions induce continuous phase transitions toward mixed states involving both singly 
and multiply-quantized vortex states. A central result is the qualitative (and experimentally 
observable) difference between power-law and hard-wall confinement: In hard-wall traps, the 
leading instability always involves states with nonzero density at the trap center, whereas 
in power-law traps the density vanishes as the rotation frequency increases. The two different
types of confinement give rise to scaling properties in the derived phase diagrams.

\end{abstract}

\maketitle

\section{Introduction}

Since the realization of Bose-Einstein condensation in cold atomic systems, their 
rotational properties have been the subject of extensive experimental and theoretical 
investigation, due to their underlying superfluid nature. The rotational response of 
these condensates depends sensitively on the form of the trapping potential, which 
experimentalists can tune with high precision. While early experiments were 
predominantly performed in harmonic traps, see, e.g., Refs.\,\cite{b0,b00,b1,b2,b3,
b4,b5,b6}, more recent advancements have enabled the realization of diverse geometries, 
including anharmonic \cite{Bretin2004,Perrin,Perrin2}, toroidal/annular \cite{c1,c2,c3,
c4,c5,c6,c7,c8,c9}, and hard-wall potentials \cite{Gaunt2013, Navon2021}.

Harmonic confinement imposes an upper bound, since the rotational frequency of the trap 
$\Omega$ cannot exceed the trap frequency $\omega$. For $\Omega = \omega$ the centrifugal 
potential (which scales quadratically with the distance from the center), $\rho$, exactly 
cancels the harmonic potential. However, in anharmonic, or hard-wall traps, the confining 
potential rises faster than quadratically with $\rho$, allowing access to regimes of much 
higher values of $\Omega$, see, e.g., the review article of Ref.\,\cite{Fetter2009} and 
Refs.\,\cite{e1,e2,e3,e4,e5,e6,e7,e8,e9,e10}. This scenario closely resembles the 
traditional problem of liquid Helium-IV rotating in a bucket. More recent
studies \cite{Machado, Molignini} have considered the stability and dynamics of quantum 
vortices in a flat potential and in a quadratic-plus-quartic trap, using beyond–mean-field 
approaches.

Motivated by these experimental advances, the present study examines the rotational 
response of a quasi-two-dimensional Bose-Einstein condensate trapped in either a tunable 
power-law potential \cite{AR} -- treating the power-law index as a parameter 
-- or a hard-wall potential. We focus on two experimentally tunable parameters: the rotational 
frequency of the trap, $\Omega$, and the coupling constant, $g$, which can be controlled via 
Feshbach resonances \cite{Fesh}. Previous studies have considered vortex stability in 
quadratic-plus-quartic traps \cite{e1,e2,e3,e4,e5,e6,e7,e8,e9,e10}. Still, the qualitative 
difference between power-law and hard-wall confinement in the instability mechanism has not 
been systematically analysed.

In both anharmonic, and hard-wall potentials, the single-particle energy 
$\epsilon_m$ exhibits a positive curvature as a function of the single-particle angular momentum 
$m \hbar$, since in both cases the potential goes to infinity faster than quadratically with 
$\rho$ (as opposed to the case of a harmonic trap, where we have a linear dependence on $m$). 
Therefore, the single-particle energy in the rotating-frame $\epsilon_m^{\rm rot} = \epsilon_m 
- m \hbar \Omega$ has a unique minimum at a specific value $m_0$, for some fixed value of
$\Omega$. As a result, for zero, or weak repulsive interatomic interactions, the many-body 
ground state is dominated by occupation of a single angular-momentum state $m_0$, which 
corresponds to a multiply-quantized vortex state. As the angular frequency of the trap 
$\Omega$ increases, transitions between states of different circulation occur discontinuously. 
Conversely, as the interaction strength $g$ increases, multiple quantization becomes 
energetically unfavorable -- see, e.g., Ref.\,\cite{Kavoulakis2000} -- resulting in 
continuous phase transitions. In these phase transitions, a multiply-quantized vortex 
state becomes unstable against a ``mixed" state, containing both singly and 
multiply-quantized vortices. 

In the results that follow below we provide a detailed and rather general 
discussion of the main instability channel of multiply-quantized vortex states, as the
interaction strength increases, or the angular frequency of the trap is varied. We also
compare the vortex-splitting behaviour, in power-law versus hard-wall confinement. 
Furthermore, we demonstrate that the phase diagrams possess a rich structure, and 
exhibit universal scaling features for the different confining potentials. Unlike 
earlier studies focusing on quadratic-plus-quartic traps, the present work systematically 
contrasts smooth power-law confinement with sharp hard-wall confinement and identifies 
the qualitative difference in the leading instability mechanism. One of the main results 
of the present work is that a hard-wall trap always leads to an instability of 
multiply-quantized vortex states to states with a non-zero density at the trap center
(at least within the approximations which are made). On the other hand, power-law traps 
exhibit a vanishing central density as the rotational frequency increases, even within
the approximations that we adopt here.

The paper is organized as follows. Section II introduces the theoretical model. In 
Sec.\,III, we analyse the discontinuous transitions driven by increasing $\Omega$, 
for weak interactions. In Sec.\,IV we examine the continuous phase transitions driven 
by increasing coupling. Universal features of the derived phase diagrams are discussed 
in Sec.\,V, and the validity of our assumptions is examined in Sec.\,VI. Section VII 
explores how the trap parameters affect the phase diagram and the density and phase 
profiles of the order parameter. Finally, we comment on the experimental relevance of 
our results in Sec.\,VIII and summarize our main conclusions in Sec.\,IX.
  
\section{Model}

The Hamiltonian that we consider is 
\begin{equation}
  {\hat H} = \sum_i \left( - \frac {\hbar^2} {2 M} {\nabla_i^2} + V({\bf r}_i) \right) 
  + \frac 1 2 {U_0} \sum_{i \neq j} \delta({\bf r}_i - {\bf r}_j).
\end{equation}
Here $M$ is the atom mass and $U_0$ is the matrix element for elastic atom-atom collisions, 
with $U_0 = 4 \pi \hbar^2 a/M$, where $a$ is the scattering length for elastic s-wave 
atom-atom collisions. The trapping potential $V({\bf r})$ is assumed to have the form
$V({\bf r}) = V_{\perp}(\rho, \theta) + V_z(z)$, where $\rho, \theta$, and $z$ are the 
usual polar coordinates. Here $V_z(z)$ is assumed to be very strong along our $z$ axis, 
which is the axis of rotation. As a result, the motion of the atoms along this direction
is frozen, and the order parameter, $\Phi(\rho, \theta, z)$ has a product form,
\begin{equation}
  \Phi(\rho, \theta, z) = \frac 1 {\sqrt Z} \Psi(\rho, \theta),
  \label{orpa}
\end{equation}
where $Z$ is the width of the condensate along the $z$ axis. In the expression above 
we assumed for simplicity that the density along this direction is constant.

From Eq.\,(\ref{orpa}) we find that the interaction energy per particle is
\begin{eqnarray}
  \frac {E_{\rm int}} N &=& \frac 1 2 N U_0 \int |\Phi(\rho, \theta, z)|^4 \, d^3 r 
\nonumber \\
 &=&  2 \pi g \frac {\hbar^2} M \int |\Psi(\rho, \theta)|^4 \, d^2 \rho,
 \label{eqintt}
\end{eqnarray} 
where $\sigma = N/Z$ is the density per unit length, with $N$ being the atom number
and $g$ the dimensionless quantity $\sigma a$.

We are interested in two types of (axially-symmetric) trapping potentials, namely 
power-law traps, 
\begin{equation}
 V_{\perp}(\rho) = \frac 1 2 M \omega^2 \rho^2 
  \left[ 1 + \lambda \left( \frac {\rho} {a_0} \right)^{2p-2} \right],
  \label{pot1}
\end{equation} 
with $p > 2$, and hard-wall potentials,
\begin{equation}
V_{\perp}(\rho) = 
\begin{cases} 
      0 & \text{if} \,\, \rho < R_0 \\
      \infty & \text{if} \,\, \rho \ge R_0. 
\end{cases}
\label{pot2}
\end{equation}
In Eq.\,(\ref{pot1}) $\omega$ is the frequency of the harmonic potential that acts on 
the $xy$ plane and $a_0=[\hbar/(M \omega)]^{1/2}$ is the oscillator length. Also, 
$\lambda$ is a small dimensionless parameter, $\lambda \ll 1$, which gives the 
``strength" of the anharmonic part of the trapping potential. Finally, in Eq.\,(\ref{pot2})
$R_0$ denotes the physical hard-wall radius used to nondimensionalize the problem. 
The parameter $R$ that we also use below is the dimensionless radius, expressed in units of 
$R_0$.

From now on we set $\hbar = M = \omega = 1$ in the problem for the anharmonic potential 
and $\hbar = M = R_0 = 1$ for the hard-wall potential.

In both cases of power-law traps and in the hard-wall potential, the
single-particle energy spectrum is discrete and it consists of various energy levels, 
$\epsilon_{n_r, m}$, where $n_r$ is the number of radial nodes, and $m$ is the 
single-particle angular momentum $m$, with the corresponding eigenfunctions denoted as
$\psi_{n_r, m}$. 

The order parameter we are interested in may be expanded in the infinite set of the
single-particle basis states, $\Psi(\rho, \theta) = \sum_{n_r, m} c_{n_r, m} \psi_{n_r, m}$. 
However, in the present study we have chosen to work in the limit of ``weak" interactions. 
Since we are interested in a rotating condensate, $m$ has to vary over a range which 
depends on the value of $\Omega$ (since we work at fixed $\Omega$), and on the interaction
strength. On the other hand, the fact that the interaction is assumed to be ``weak", we may 
ignore the states with nonzero radial nodes, $n_r > 0$, and assume that $\Psi(\rho, \theta) 
= \sum_{m} c_{m} \psi_{n_r=0, m}$. From now on, we denote $\psi_{n_r=0, m}$ as 
$\psi_{m}^{\rm LLL}$ for the anharmonic potential and $\psi_{m}^{\rm HD}$ for the hard-wall 
potential, while the corresponding eigenenergies $\epsilon_{n_r, m}$ are denoted as 
$\epsilon_{m}$. The assumption of weak interactions is examined further in Sec.\,VI.
 
\section{Phase diagram - Discontinuous transitions}

We begin by analysing the phase diagram in the weak-interaction regime. In general 
one has to solve the single-particle problem for every form of the trapping potential. 
The resulting eigenvalues will be characterized by two quantum numbers, namely the 
radial quantum number $n$ and the one that corresponds to the angular momentum, $m$. 

Regarding the anharmonic potential, we will make two simplifying assumptions, which 
will be examined later, in Sec.\,VI. The first one is that the value of $\lambda$ 
in Eq.\,(\ref{pot1}), i.e., the strength of the anharmonic term in the potential is 
small and thus we will treat it perturbatively, working with the eigenstates of the 
harmonic potential (in two dimensions). The second assumption is that the interaction 
is weak, in the sense that the typical energy associated with the interaction energy 
is lower than the energy difference between the lowest-energy eigenvalues $\epsilon_m$ 
with no radial nodes and the ones with one radial node (obviously, all the other ones 
will have an even larger energy). In the case of the hard-wall potential, the only 
assumption that we make is the last one. 

Throughout this work, we denote by $\psi_m(\rho, \theta)$ the single-particle 
eigenstates of the non-interacting problem, with angular momentum $m$. When needed, 
we distinguish between the lowest-Landau-level states $\psi_m^{\rm LLL}$ for power-law 
traps and the hard-wall eigenstates $\psi_m^{\rm HD}$. The two-dimensional condensate 
order parameter, which is denoted by $\Psi(\rho, \theta)$ [see Eq.\,(\ref{orpa})], is 
expressed as a superposition of the single-particle states.

As mentioned above, since all the potentials that we consider grow more rapidly 
than the harmonic, the single-particle energy $\epsilon_m$ has a positive curvature 
as a function of $m$. If we go to the rotating frame, the energy (for zero coupling) 
\begin{equation}
  \epsilon_m^{\rm rot} = \epsilon_m - m \Omega 
\end{equation}
also has a positive curvature. For $\Omega$ equal to zero, or small enough, the minimum 
of $\epsilon_m^{\rm rot}$ occurs at $m_0 = 0$. As $\Omega$ increases, eventually the 
minimum of $\epsilon_m^{\rm rot}$ shifts to $m_0 = 1$, and so on.  As a result, as 
$\Omega$ increases, there are discontinuous phase transitions between multiply-quantized 
vortex states of different circulation, and the order parameter is of the form
\begin{equation}
   \Psi(\rho, \theta) = f(\rho) \, e^{i m_0 \theta}.
   \label{op1}
\end{equation} 
While the above remarks are general and hold for both types of trapping potential, we 
distinguish between the two kinds of trapping potential in order to get some explicit 
results.

\subsection{Power-law traps}

Let us start with the anharmonic potential of Eq.\,(\ref{pot1}). The assumptions 
that we mentioned above, namely that the anharmonic part of the trapping potential 
is weak, and also that the interaction is weak, allow us to work with the lowest-Landau 
level states of the harmonic potential,
\begin{equation}
  \psi_{m}^{\rm LLL}(\rho, \theta) = \frac 1 {\sqrt{\pi m!}} \rho^m e^{-\rho^2/2} e^{i m \theta},
\end{equation}
with an eigenenergy which is equal to $\epsilon_m = m + 1$ (for $m \ge 0$). According 
to first-order perturbation theory, the anharmonic part of the trapping potential 
$\Delta V$ contributes a term in the single-particle energy which is 
\begin{equation}
   \langle \psi_m^{\rm LLL} | \Delta V | \psi_m^{\rm LLL} \rangle 
   = \frac {\lambda} 2 \frac {(m+p)!} {m!}.
\end{equation}
Finally, the interaction energy per particle is, according to Eq.\,(\ref{eqintt}),
\begin{equation}
  \frac {E_{{\rm int},m}} N = g \frac {(2 m)!} {2^{2m} (m!)^2}.
\end{equation}
Collecting the above results, we get that the energy per particle in the rotating 
frame is
\begin{eqnarray}
  \frac {E_m^{\rm rot}} N &=& \epsilon_m - m \Omega + 
  \langle \psi_m^{\rm LLL} | \Delta V | \psi_m^{\rm LLL} \rangle
  + \frac {E_{{\rm int},m}} N
  \nonumber \\
  &=& 1 + m (1 - \Omega) + \frac {\lambda} 2 \frac {(m+p)!} {m!}
  + g  \frac {(2 m)!} {2^{2m} (m!)^2}.
  \nonumber \\
  \label{erottot}
\end{eqnarray}
The transition between multiply-quantized vortex states with circulation 
$m$ and $m+1$ occurs when $E_{m+1}^{\rm rot} = E_{m}^{\rm rot}$, which determines the 
critical rotation frequency $\Omega_{m, m+1}$. From Eq.\,(\ref{erottot}) it follows that
\begin{eqnarray}
  \Omega_{m, m+1} = 1 + \frac{\lambda}{2} 
  \left[ \frac{(m+1+p)!}{(m+1)!} - \frac{(m+p)!}{m!} \right] +
\nonumber \\
+ g \left( \frac {(2 m + 2)!} {2^{2m+2} [(m+1)!]^2} 
- \frac {(2 m)!} {2^{2m} (m!)^2} \right).
 \label{eqq}
\end{eqnarray}
Equation (\ref{eqq}) defines straight lines on the $\Omega - g$ plane, which 
constitute part of the phase diagram we are looking for; see the solid (red) 
lines, in Figs.\,1 and 3, which are given by Eq.\,(\ref{eqq}), and correspond to discontinuous 
transitions. The dotted (black) curves shown in the same figures correspond to continuous 
transitions and are analysed in Sec.\,IV. Clearly, the state with some given $m = m_0$ is 
stable in the range $\Omega_{m_0-1,m_0} < \Omega < \Omega_{m_0, m_0+1}$.

\begin{figure}[t]
\includegraphics[width=7cm,height=6cm,angle=0]{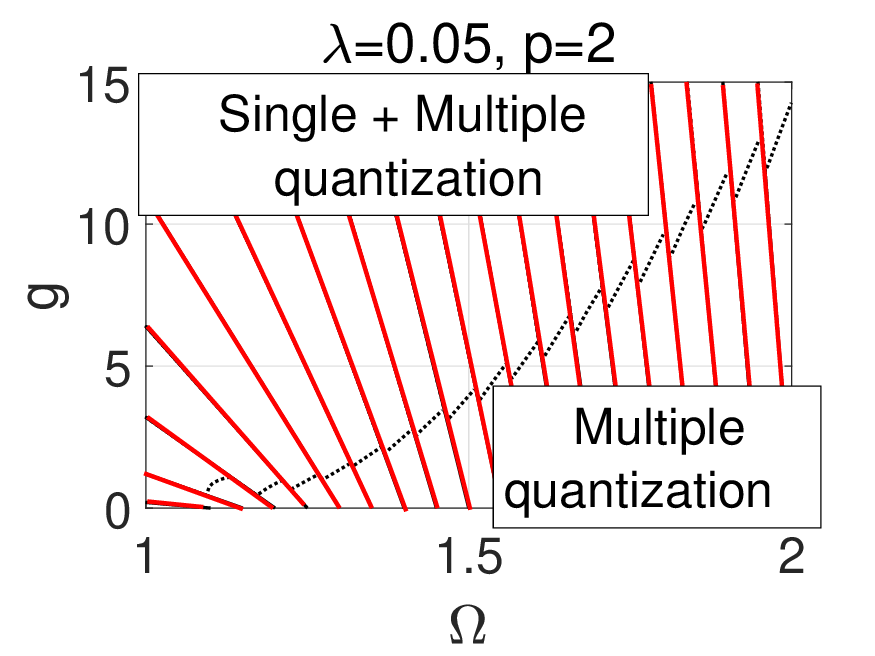}
\includegraphics[width=7cm,height=6cm,angle=0]{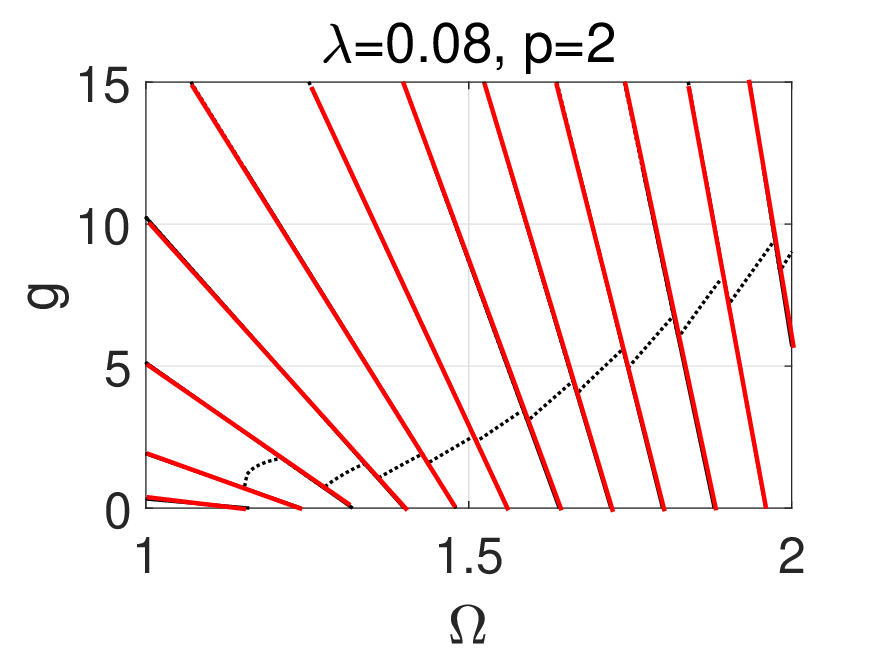}
\caption{The phase diagram, where on the $x$ axis is the rotational frequency
of the trap $\Omega$ (in units of $\omega$) and on the $y$ axis is the (dimensionless) 
coupling $g$. Here we have a power-law trapping potential, with $p = 2$, $\lambda = 0.05$ 
(higher) and $\lambda = 0.08$ (lower). The solid (red) lines, which are 
given by Eq.\,(\ref{eqq}), give the discontinuous transitions, while the dotted (black) curves 
give the continuous ones. The dotted (black) curves separate the phase diagram into two parts. 
In the lower part the order parameter corresponds to vortices of multiple quantization. In the 
upper part, the order parameter corresponds to a ``mixed" state of single and multiple 
quantization. In the region of multiple quantization, each time one crosses a solid (red) 
line, the value of $m_0$ increases by one unit, starting from $m_0 = 0$.}
\end{figure}

\begin{figure}[t]
\includegraphics[width=9.5cm,height=8.2cm,angle=0]{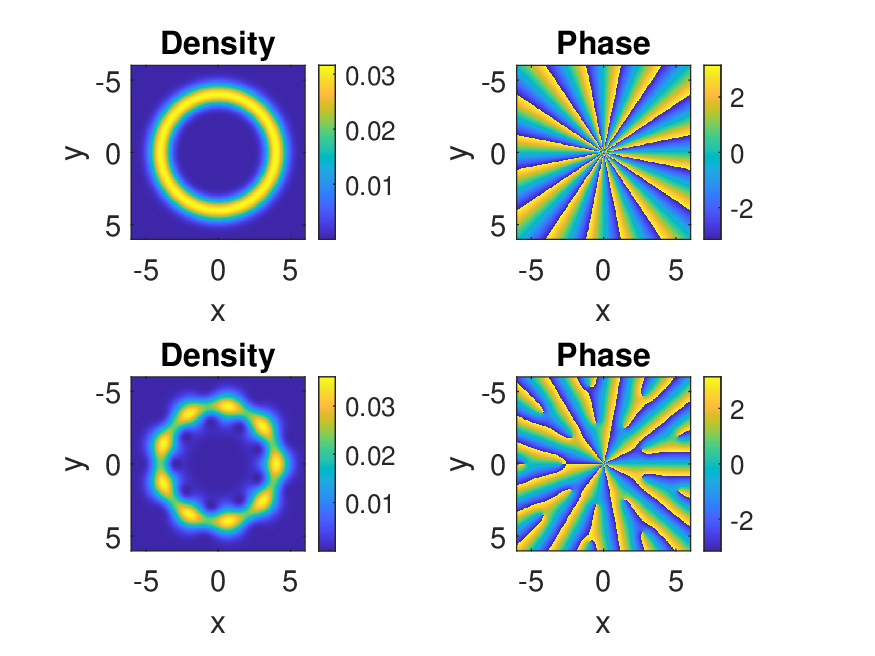}
\caption{(Colour online). The density (left) and the phase (right) of the 
order parameter -- with $p=2$ and $\lambda = 0.08$ -- for the case of 
a multiply-quantized vortex state in a power-law trap, with $m_0 = 16$ 
(higher), towards the instability involving the states with $(m_0-q, m_0, m_0+q) 
= (7, 16, 25)$ (lower). Here the axes are measured in units of $a_0$, and the 
density in units of $a_0^{-2}$.}
\end{figure}

\begin{figure}[h]
\includegraphics[width=7cm,height=6cm,angle=0]{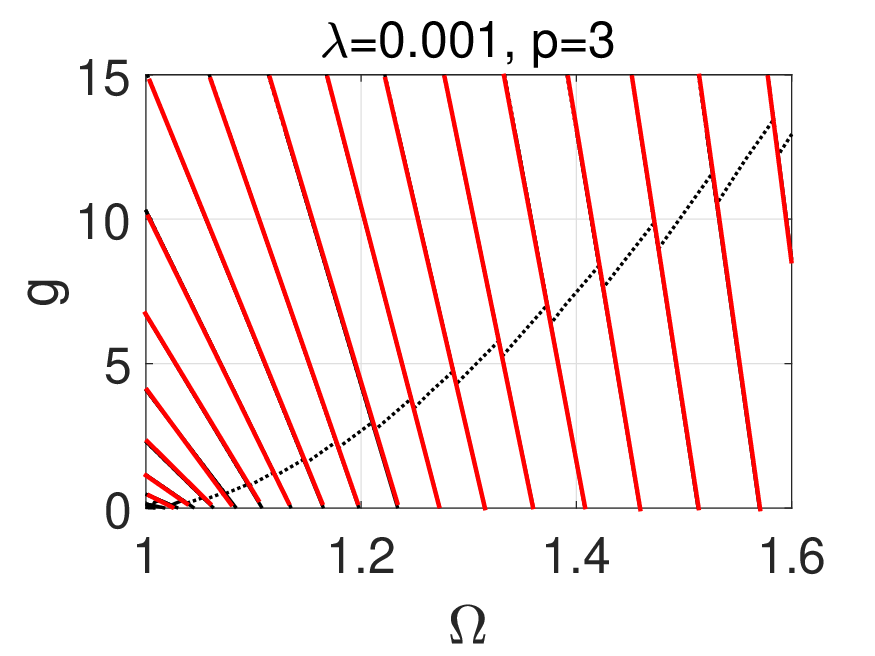}
\includegraphics[width=7cm,height=6cm,angle=0]{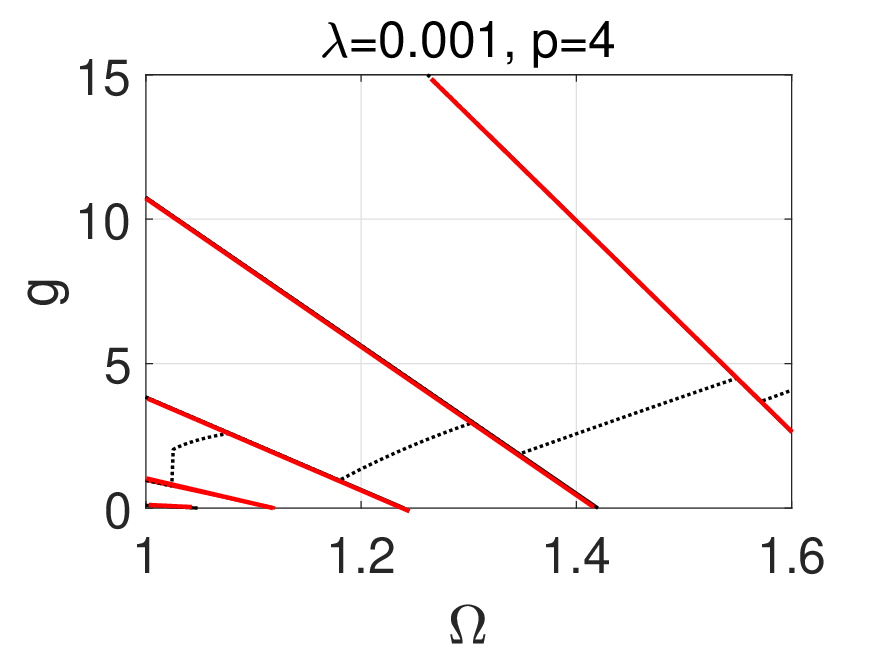}
\caption{The phase diagram, where on the $x$ axis is the rotational frequency
of the trap $\Omega$ (in units of $\omega$) and on the $y$ axis is the (dimensionless)
coupling $g$. Here we have a power-law trapping potential, with $\lambda = 0.001$, $p = 3$ 
(higher) and $p = 4$ (lower). The solid (red) lines, which are given by  
Eq.\,(\ref{eqq}), give the discontinuous transitions, while the dotted (black) curves give the 
continuous ones. The dotted (black) curves separate the phase diagram into two parts. In the lower
part the order parameter corresponds to vortices of multiple quantization. In the upper part, 
the order parameter corresponds to a ``mixed" state of single and multiple quantization.
In the region of multiple quantization, each time one crosses a solid (red) line, the value 
of $m_0$ increases by one unit, starting from $m_0 = 0$.}
\end{figure}

\begin{figure}[h]
\includegraphics[width=9.5cm,height=8.2cm,angle=0]{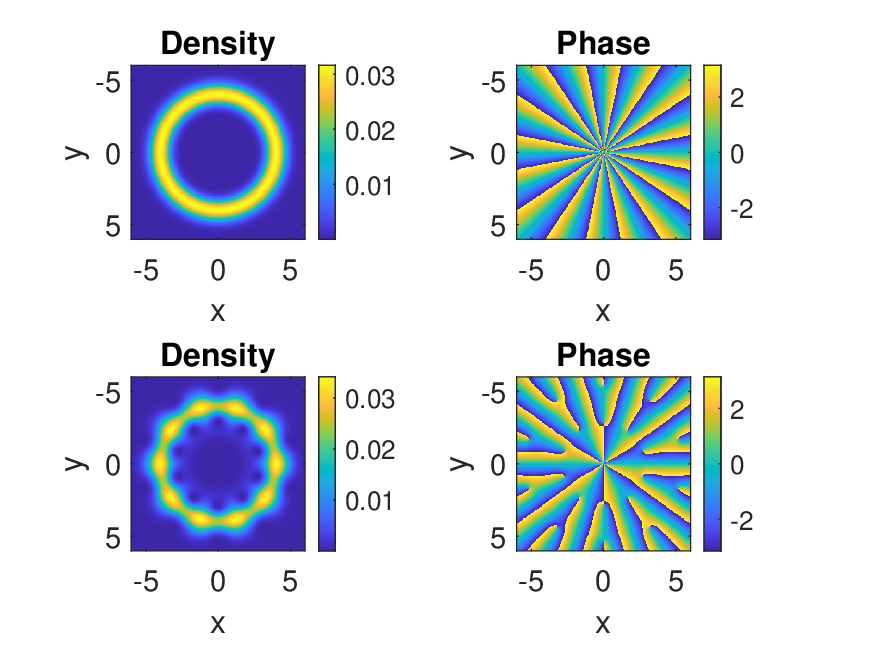}
\caption{(Colour online). The density (left) and the phase (right) of the 
order parameter -- with $p=3$ and $\lambda = 0.001$ -- for the case of 
a multiply-quantized vortex state in a power-law trap, with $m_0 = 16$ 
(higher), towards the instability involving the states with $(m_0-q, m_0, m_0+q) 
= (6, 16, 26)$ (lower). Here the axes are measured in units of $a_0$, and the 
density in units of $a_0^{-2}$.}
\end{figure}

\begin{figure}[h]
\includegraphics[width=9.5cm,height=8.2cm,angle=0]{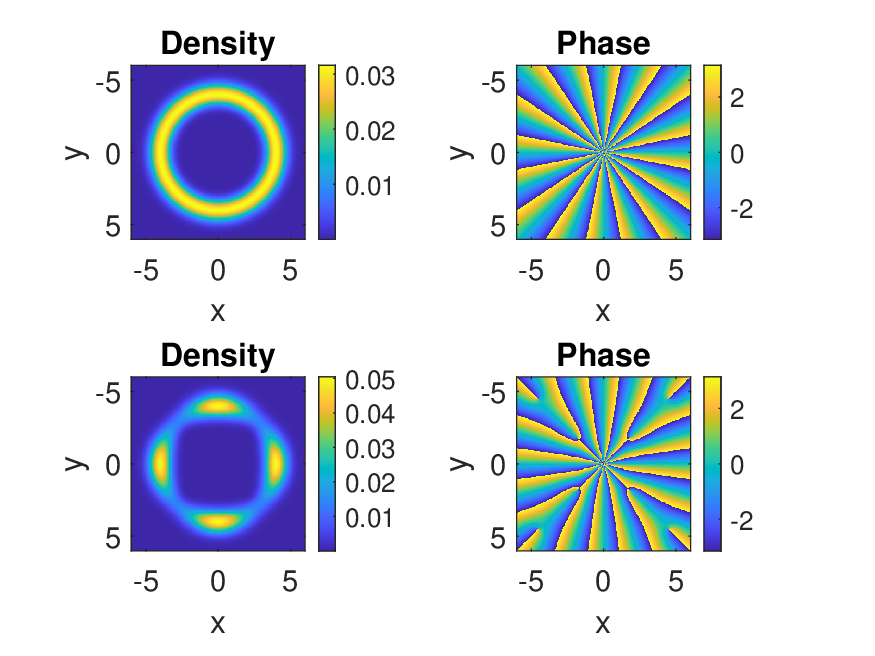}
\caption{(Colour online). The density (left) and the phase (right) of the 
order parameter, for the case of multiply-quantized vortex state in a 
power-law trap, with $p=4$ and $\lambda = 0.001$, $m_0 = 16$ (higher), 
towards the instability involving the states with $(m_0-q, m_0, m_0+q)
=(12, 16, 20)$ (lower). Here the axes are measured in units of $a_0$, 
and the density in units of $a_0^{-2}$.}
\end{figure}

We now turn to the limit of rapid rotation, where $m \gg 1$. Regarding the term 
that involves $\lambda$,
\begin{equation}
   \frac{\lambda}{2} 
  \left[ \frac{(m+1+p)!}{(m+1)!} - \frac{(m+p)!}{m!} \right] \approx 
  \frac {\lambda p} 2 m^{p-1}.
\end{equation}
Also, regarding the interaction term, since
\begin{equation}
   \frac {E_{{\rm int},m}} N 
   \approx \frac g {\sqrt{\pi m}},
\end{equation}
one finds that
\begin{equation}
   \frac 1 N [{E_{{\rm int},m+1}} - {E_{{\rm int},m}}] 
   \approx - \frac g {2 \sqrt{\pi}} m^{-3/2}.
\end{equation}
Collecting these results, Eq.\,(\ref{eqq}) implies that
\begin{eqnarray}
  \Omega_{m, m+1} \approx 1 + \frac{\lambda p}{2} m^{p-1} - \frac{g}{2 \sqrt \pi} m^{-3/2},
\label{neqqq}
\end{eqnarray}
or
\begin{eqnarray}
  g \approx 2 \sqrt \pi m^{3/2} 
   \left( 1 + \frac{\lambda p}{2} m^{p-1} - \Omega_{m, m+1} \right).
\label{rvsom1}
\end{eqnarray}
According to Eq.\,(\ref{eqq}) the slope of $g = g(\Omega)$ is negative, being
proportional to $m^{3/2}$, tending to minus infinity for large $m$, i.e., the
corresponding lines tend to become vertical. Also, the spacing between successive
lines for $g = 0$ is 
\begin{eqnarray}
   \Delta \Omega_m &\equiv& \Omega_{m+1, m+2} - \Omega_{m, m+1} 
   \nonumber \\
   &\approx& \frac {\lambda p} 2 [(m+1)^{p-1} - m^{p-1}]
  \nonumber \\
  &\approx& \frac {\lambda p (p-1)} {2} m^{p-2}.
\label{rvsom11}
\end{eqnarray}
Therefore, for $p=2$, $\Delta \Omega_{m, m+1}$ is independent of $m$, being equal to 
$\lambda$, while for any $p>2$, $\Delta \Omega_{m, m+1}$ increases with increasing $m$.

\subsection{Hard-wall traps}

We now turn to the case of a hard-wall potential of some radius $R$, which is measured
in units of $R_0$. The single-particle problem is
\begin{eqnarray}
  - \frac {1} {2} \nabla^2 \psi^{\rm HD}(\rho, \theta) = \epsilon \psi^{\rm HD}(\rho, \theta),
\end{eqnarray}
with the boundary condition $\psi^{\rm HD}(\rho = R, \theta) = 0$. The above equation takes 
the form
\begin{eqnarray}
  \nabla^2 \psi^{\rm HD}(\rho, \theta) + k^2 \psi^{\rm HD}(\rho, \theta) = 0,
\end{eqnarray}
with $k^2 = 2 \epsilon$. From the expression of the Laplacian in two dimensions,
\begin{eqnarray}
  \psi_{\rho \rho}^{\rm HD}(\rho, \theta) + \frac 1 {\rho} \psi_{\rho}^{\rm HD}(\rho, \theta)
  + \frac 1 {\rho^2} \psi_{\theta \theta}^{\rm HD}(\rho, \theta) + k^2 \psi^{\rm HD}(\rho, \theta) = 0.
\nonumber \\
\end{eqnarray}
With the assumption that $\psi^{\rm HD}(\rho, \theta) = f(\rho) e^{i m \theta}$,
with $m$ being an integer that corresponds to the angular momentum,
\begin{eqnarray}
  {f}_{\rho \rho}(\rho) + \frac 1 {\rho} {f}_{\rho}(\rho)
  - \frac {m^2} {\rho^2} {f}(\rho) + k^2 {f}(\rho) = 0.
\end{eqnarray}
The functions $f(\rho)$ are Bessel functions $J_{m}(k \rho)$, while the eigenfunctions 
(which are characterized by two quantum numbers, $n$ and $m$), are 
\begin{equation}
    \psi_{n,m}^{\rm HD}(\rho, \theta) = N_{n,m} J_m(k_{n,m} \rho) e^{i m \theta},
\end{equation}
where $N_{n,m}$ are the normalization constants. The boundary condition due to the 
assumption of a hard-wall potential $J_m(k_{n,m} R) = 0$ implies that $k_{n,m} R = 
\alpha_{n,m}$, where $\alpha_{n,m}$ are the roots of $J_m(x)$, and therefore $k_{n,m} 
= \alpha_{n,m}/R$. Therefore, the single-particle energy spectrum is
\begin{equation}
  \epsilon_{n,m} = \frac {\alpha_{n,m}^2} {2 R^2}. 
\end{equation}
Since we make again the assumption of weak interactions, we restrict ourselves
to the roots with $n=0$. Therefore, we work with the single-particle states 
$\psi_{0,m}^{\rm HD}(\rho, \theta)$, with the corresponding eigenenergies denoted 
as $\epsilon_{0,m}$. From now on, we drop the index ``0" and use the notation 
$\psi_{m}^{\rm HD}(\rho, \theta)$ for the eigenstates and $\epsilon_m$ for the 
eigenvalues which correspond to the roots $\alpha_{0,m}$.

\begin{figure}[h]
\includegraphics[width=7cm,height=5cm,angle=0]{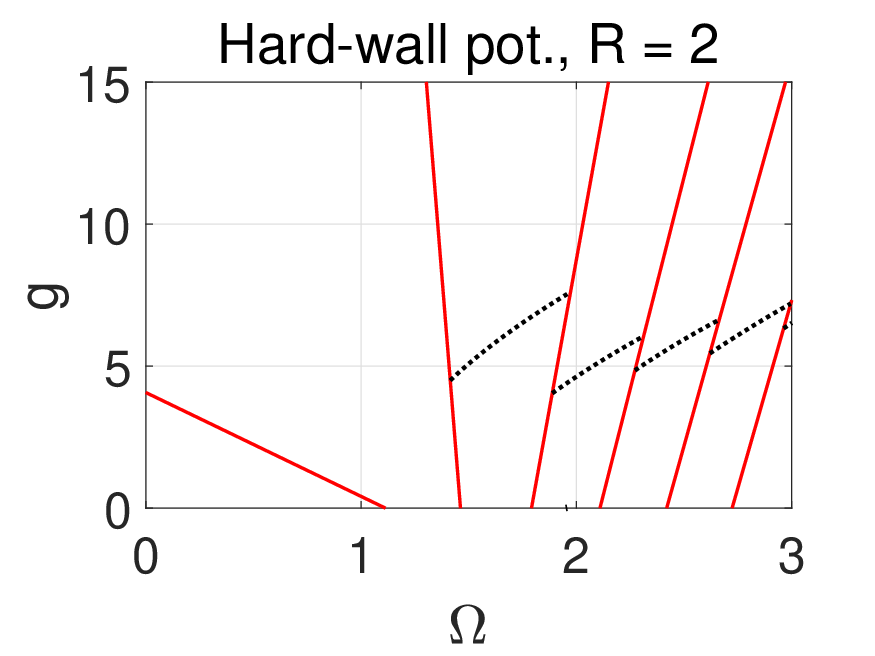}
\includegraphics[width=7cm,height=5cm,angle=0]{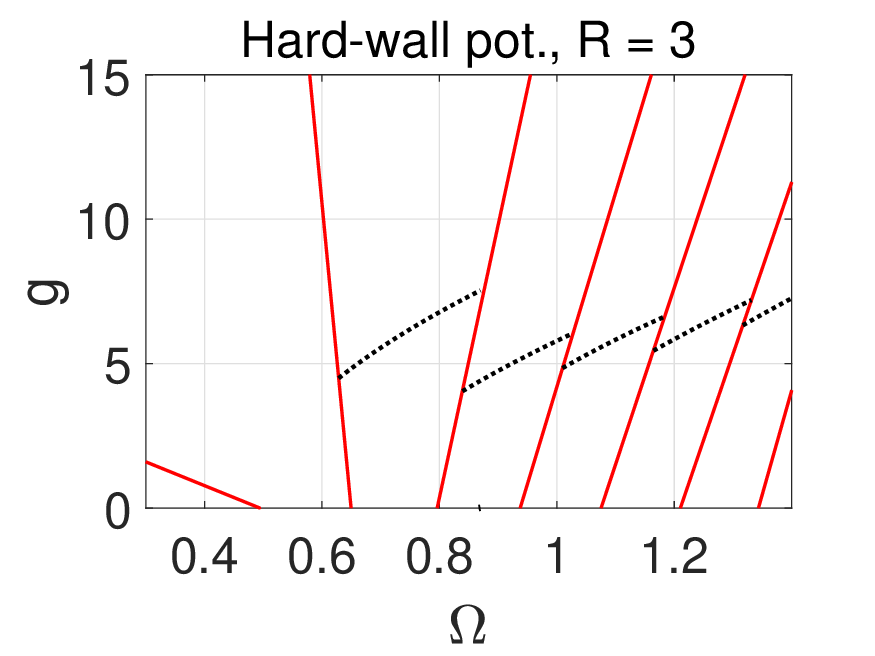}
\includegraphics[width=7cm,height=5cm,angle=0]{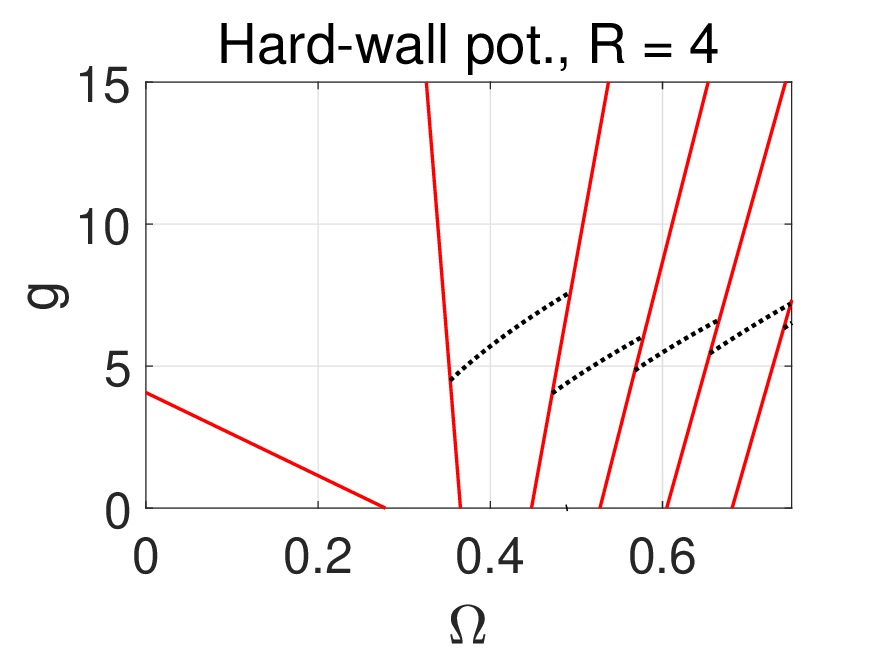}
\caption{The phase diagram, where on the $x$ axis is the rotational frequency
of the trap $\Omega$ (in units of $\omega$) and on the $y$ axis is the (dimensionless)
coupling $g$. Here we have a hard-wall trapping potential, with $R = 2$ (higher), 
$R = 3$ (middle), and $R = 4$ (lower). Here $R$ is measured in units of $R_0$.
The solid (red) lines, which are given by Eq.\,(\ref{neqq}), 
give the discontinuous transitions, while the dotted (black) curves give the 
continuous ones. The dotted (black) curves separate the phase diagram into two parts. 
In the lower part the order parameter corresponds to vortices of multiple quantization. 
In the upper part, the order parameter corresponds to a ``mixed" state. 
In the region of multiple quantization, each time one crosses a solid (red) line, 
the value of $m_0$ increases by one unit, starting from $m_0 = 0$.}
\end{figure}

\begin{figure}[h]
\includegraphics[width=9.5cm,height=8.2cm,angle=0]{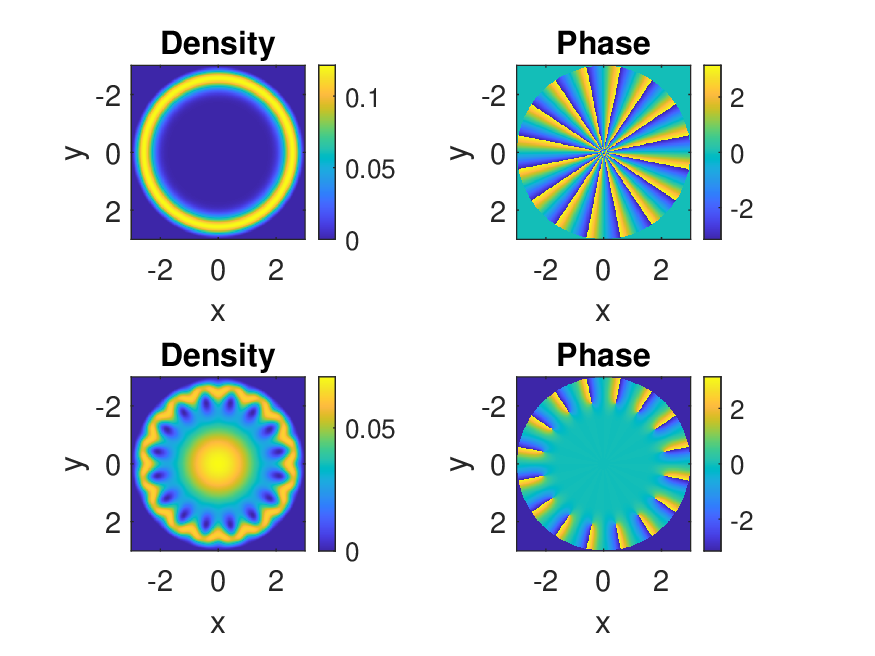}
\caption{(Colour online). The density (left) and the phase (right) of the 
order parameter, in a hard-wall potential, for the case of a multiply-quantized 
vortex state, with $m_0 = 16$ (higher), towards the instability involving the 
states with $(m_0-q, m_0, m_0+q) = (0, 16, 32)$ (lower). Here $R = 3$ in units 
of $R_0$, while the axes are measured in units of $R_0$, and the density in units 
of $R_0^{-2}$.}
\end{figure}

The total energy per particle in the rotating frame is 
\begin{eqnarray}
  \frac {E_m^{\rm rot}} N = \frac {\alpha_{0,m}^2} {2 R^2} - m \Omega 
  + 2 \pi g \int |\psi_m^{\rm HD}|^4 \, d^2 \rho.
  \label{erottot2}
\end{eqnarray}
The transition between states with circulation $m$ and $m+1$ occurs when 
$E_{m+1}^{\rm rot} = E_{m}^{\rm rot}$. From Eq.\,(\ref{erottot2}) it follows that
\begin{eqnarray}
  \Omega_{m, m+1} = \frac {\alpha_{0,m+1}^2 - \alpha_{0,m}^2} {2 R^2} +
\nonumber \\
  + 2 \pi g \int [|\psi_{m+1}^{\rm HD}|^4 - |\psi_m^{\rm HD}|^4] \, d^2 \rho.
  \label{neqq}
\end{eqnarray}
Again, the last equation defines straight lines on the $\Omega - g$ plane; see the 
lines in Fig.\,6.

In the limit of rapid rotation, where $m \gg 1$, since
\begin{equation}
  \alpha_{0,m} = m + {\cal O} (m^{1/3}),
\end{equation}
therefore 
\begin{equation}
  \frac {\alpha_{0,m+1}^2 - \alpha_{0,m}^2} {2 R^2} \approx \frac {m} {R^2}.
\end{equation}
Regarding the interaction term, we find numerically that
\begin{equation}
  \frac {E_{{\rm int},m}} N 
  =  2 \pi g \int |\psi_m^{\rm HD}|^4 \, d^2 \rho \approx \frac {2 \pi g C} {R^2} {\sqrt{m}}, 
\end{equation}
where $C \approx 0.308$. Therefore, 
\begin{eqnarray}
  \frac 1 N [{E_{{\rm int},m+1}} - {E_{{\rm int},m}}] 
  &=& 2 \pi g \int [|\psi_{m+1}^{\rm HD}|^4 - |\psi_m^{\rm HD}|^4] \, d^2 \rho 
  \nonumber \\
  &\approx& \frac {\pi g C} {R^2} \frac 1 {\sqrt{m}}. 
\end{eqnarray}
From the equations above it follows that
\begin{eqnarray}
  \Omega_{m, m+1} \approx \frac {m} {R^2} + \frac {g \pi C} {R^2} \frac 1 {\sqrt{m}},
  \label{qqee}
\end{eqnarray}
or
\begin{eqnarray}
  g \approx R^2 \frac {\sqrt m} {\pi C} \left( \Omega_{m, m+1} - \frac {m} {R^2} \right).
\label{rvsom12}
\end{eqnarray}
According to Eq.\,(\ref{rvsom12}) the slope of $g = g(\Omega)$ is positive, being
proportional to $m^{1/2}$, tending to plus infinity for large $m$, i.e., the
corresponding lines tend to become vertical. Also, the spacing between successive
lines for $g = 0$ is 
\begin{eqnarray}
   \Delta \Omega_m 
   \equiv \Omega_{m+1, m+2} - \Omega_{m, m+1} \approx \frac 1 {R^2}.
\label{rvsom122} 
\end{eqnarray}
Therefore, $\Delta \Omega_{m, m+1}$ becomes independent of $m$, for large $m$.

In conclusion, at least for large values of $m$, the function $g=g(\Omega)$
that corresponds to the discontinuous phase transitions, scales as $-m^{3/2} \Omega$ for
the power-law traps and as $+m^{1/2} \Omega$ for the hard-wall potential. This highlights 
a fundamental physical distinction between the two types of confinement.

\section{Phase diagram - Continuous transitions}

We now turn to the problem of continuous phase transitions. As discussed above, as the 
interaction strength increases, multiply-quantized vortex states become energetically 
unfavourable and undergo continuous phase transitions toward states containing in 
general vortices of single and multiple quantization. Due to angular momentum 
conservation, a multiply-quantized vortex state $\psi_{m_0}$ becomes unstable 
toward the triplet of states \cite{e9,e10,Kavoulakis2000}
\begin{equation}
   \Psi(\rho, \theta) = c_{m_0-q} \psi_{m_0-q} + c_{m_0} \psi_{m_0} + c_{m_0+q} \psi_{m_0+q},
\label{varphi}
\end{equation} 
with $1 \le q \le m_0$, where $\psi_{m}(\rho, \theta)$ are the lowest-energy single-particle 
eigenstates of the non-interacting problem (i.e., either $\psi_m^{\rm LLL}$ for the anharmonic
potential, or $\psi_m^{\rm HD}$ for the hard-wall potential), with a corresponding eigenenergy 
$\epsilon_m$. We stress that the amplitudes $c_{m_0 \pm q}$ in the above equation 
vanish below the phase boundary and increase continuously as one crosses the phase boundary. 
As a result, we have continuous phase transitions.

In order to gain insight, we mention that the transition energy cost 
is affected by an increase in $q$. Physically, the kinetic energy cost of exciting 
$\psi_{m_0 \pm q}$ increases significantly with $q$, due to the positive curvature of 
the single-particle energy spectrum $\epsilon_m$. However, the interaction matrix 
elements that drive the instability also depend on $q$. The continuous phase boundary is 
determined by a competition where the system selects the specific value of $q$ that minimizes 
the net energy cost (balancing the kinetic energy against the interaction energy). This 
explains why only one specific value of $q$ becomes the leading instability channel for 
a given $m_0$, as summarized in Table I. 

Furthermore, the leading instability is always of the form of Eq.\,(\ref{varphi}) (i.e., 
involving three states), since we have two-body collisions. As a result, the interaction 
energy involves the off-diagonal term $c_{m-q} c_{m_0}^2 c_{m+q}$, due to angular-momentum 
conservation in the collisions. The end result is a quadratic form in the amplitudes $c_{m-q}$
and $c_{m+q}$. The ultimate goal of this approach is, for fixed $\Omega$, the identification 
of the dominant instability that corresponds to some value of $q$ which minimizes the critical 
coupling $g$.

Close to the phase boundary $|c_{m_0}| \approx 1$, while $|c_{m_0-q}|$ and $|c_{m_0+q}|$ are 
very small allowing a linearization of the energy. The energy in the rotating frame in the 
state of Eq.\,(\ref{varphi}) is (keeping only terms which are up to quadratic in 
$c_{m_0 \pm q}$),
\begin{eqnarray}
  \frac {E^{\rm rot}} N &=& \frac 1 S [\epsilon_{m_0-q}^{\rm rot} c_{m_0-q}^2 +
  \epsilon_{m_0}^{\rm rot} c_{m_0}^2 + \epsilon_{m_0+q}^{\rm rot} c_{m_0+q}^2]
\nonumber \\
&+& \frac 1 {S^2} 2 \pi g [c_{m_0}^4 \langle m_0, m_0 | m_0, m_0 \rangle +
\nonumber \\
&+& 4  c_{m_0}^2 c_{m_0 - q}^2 \langle m_0, m_0-q | m_0, m_0-q \rangle +
\nonumber \\
&+& 4  c_{m_0}^2 c_{m_0 + q}^2 \langle m_0, m_0+q | m_0, m_0+q \rangle +
\nonumber \\
&+& 4 c_{m_0}^2 c_{m_0 - q} c_{m_0 + q} \langle m_0, m_0 | m_0-q, m_0+q \rangle],
\nonumber \\
\label{quaddd}
\end{eqnarray} 
where  
\begin{equation}
  \langle m_i, m_j | m_k, m_l \rangle = \int \psi_{m_i}^* \psi_{m_j}^*
  \psi_{m_k} \psi_{m_l} \, d^2 \rho.
  \label{quad}
\end{equation}
Here normalization is enforced via $S = \sum_m c_m^2$, and the derivatives are taken 
with respect to $c_{m_0 \pm q}$, without expanding the factors $1/S$ and $1/S^2$. 
From the quadratic form of Eq.\,(\ref{quaddd}) one obtains the Hessian matrix that 
determines the stability of the state $\psi_{m_0}$, where the diagonal elements are
\begin{eqnarray}
 \frac {\partial^2 (E^{\rm rot}/N)} {\partial c_{m_i}^2} &=& 2 [\epsilon_{m_i}^{\rm rot} 
 - \epsilon_{m_0}^{\rm rot}]
 + 16 \pi g \langle m_0, m_i | m_0, m_i \rangle
 \nonumber \\ 
 &-& 8 \pi g \langle m_0, m_0 | m_0, m_0 \rangle, 
 \label{diagg}
\end{eqnarray}
and the off-diagonal are
\begin{equation}
 \frac {\partial^2 (E^{\rm rot}/N)} {\partial c_{m_i} \partial c_{m_j}} = 
 8 \pi g \langle m_0, m_0 | m_i, m_j \rangle.
\label{diaggg}
\end{equation}
The onset of an instability of the multiply-quantized state that we are looking for shows up 
when one of the eigenvalues of this matrix becomes negative. Clearly the above approach leads 
to $2 \times 2$ matrices. 

For some given value of $m_0$, one has to consider all the allowed values of $q$, 
i.e., $1 \le q \le m_0$. The more general problem will be block diagonal, with each matrix, 
describing the instability from some $\psi_{m_0}$ to the triplet $(\psi_{m_0-q}, \psi_{m_0}, 
\psi_{m_0+q})$ still being $2 \times 2$. The (leading) instability is the one that involves 
the specific value of $q$ with the smallest value of the coupling, $g$.

The above general formalism allows us to study this problem for any trapping potential. 
In the case of the (weakly) anharmonic potential the matrix elements may be evaluated 
analytically. In the case of the hard-wall potential, these are evaluated numerically.

Using the approach that we just described, we plot in Figs.\,1, 3, and 6 
the phase boundaries for the continuous phase transitions, denoted as dotted (black) 
curves for the case of the anharmonic potential (Figs.\,1 and 3), and for the hard-wall 
potential (Fig.\,6), along with the straight lines which correspond to the discontinuous 
phase transitions solid (red) lines (which were analysed in the previous 
section). The dotted (black) curves separate all these phase diagrams into 
two parts. In the lower part the order parameter is given by a ``mixed" state. 
In the upper part, the order parameter corresponds in general to a state of single and 
multiple quantization. Finally, in Figs.\,2, 4, 5, and 7 we plot the density and phase 
of the corresponding order parameter, for the multiply-quantized vortex states and the 
``mixed" ones. An analysis of the derived phase diagrams follows below, in Sec.\,VII. 

We stress that while the method we have adopted allows us to produce the 
phase diagrams of Figs.\,1, 3, and 6, on the other hand it does not give us the 
amplitudes of the order parameter in the ``mixed" states – see, e.g., Eq.\,(\ref{varphi}). 
As a result, the density and phase plots in Figs. 2, 4, 5, and 7 are representative, in 
the sense that they only illustrate the general topological characteristics, symmetries, 
and phase configurations of the order parameter. 

\section{Universality of the phase diagram}

\subsection{Power-law traps}

In the case of power-law traps, there is a scaling which makes the phase
diagram universal, as was demonstrated in Ref.\,\cite{e9}. In Eq.\,(\ref{diagg}),
\begin{eqnarray}
  \epsilon_{m_i}^{\rm rot}  - \epsilon_{m_0}^{\rm rot} = (1 - \Omega) (m_i - m_0) +
  \nonumber \\ 
  + \frac {\lambda} 2 \left( \frac {(m_i+p)!} {m_i!} - \frac {(m_0+p)!} {m_0!} \right).
\label{diagggg}
\end{eqnarray}
From Eqs.\,(\ref{diagg}), (\ref{diaggg}), and (\ref{diagggg}) it follows that the 
Hessian is linear in $1 - \Omega$, $\lambda$, and $g$. Therefore a common rescaling 
of these three parameters by a factor $\beta$ multiplies the Hessian by $\beta$. 
The vanishing of the eigenvalues is thus unaffected by this rescaling and the phase 
diagram is unaltered.

\subsection{Hard-wall traps}

In contrast to the power-law case, the hard-wall potential introduces an intrinsic 
length scale through the radius $R$, breaking the scale invariance of the problem. 
Here, since both the kinetic, as well as the interaction energies scale as $R^{-2}$, 
the phase diagram is universal when expressed in terms of $g$ and $\Omega R^2$. 

This is the reason why the top and bottom plots in Fig.\,6 are identical (up to a 
rescaling of the horizontal axis). The only difference between the two is the value 
of $R$, being 2 in the top and 4 in the bottom one. Also, the $x$ axis, i.e., the 
$\Omega$ axis, has been rescaled by a factor of 4. We stress that in the case of 
power-law traps this type of scaling is absent, due to the different form of the
trapping potential.

\begin{table}[htbp]
\centering
\caption{Instability of the multiply-quantized vortex states $\psi_{m_0}$ towards
a linear superposition involving the states $(\psi_{m_0-q}, \psi_{m_0}, \psi_{m_0+q})$. 
The first column gives the value of $m_0$, the second gives the value of $q$ for
$(p=2, \lambda = 0.05)$, the third for $(p=2, \lambda = 0.08)$, the fourth $(p=3, 
\lambda = 0.001)$ the fifth $(p=4, \lambda = 0.001)$, the last one corresponds
to a hard-wall potential, with $R = 3$.}
\label{tab:rotational_response}
\begin{tabular}{ | p{0.8cm} | p{1.1cm} | p{1.1cm} | p{1.1cm} | p{1.1cm} | p{1.1cm} |}
\hline
 & $p=2, \linebreak \lambda = 0.05$ & $p=2, \linebreak \lambda = 0.08$ & $p = 3, \linebreak \lambda = 0.001$ & $p = 4, \linebreak \lambda = 0.001$ & Hard \linebreak wall $(R=3)$ \\
\hline
$m_0$ & $q$ & $q$ & $q$ & $q$ & $q$ \\
\hline
2 & 2 & 2 & 2 & 2 & 2 \\
\hline
3 & 3 & 3 & 3 & 3 & 3 \\
\hline
4 & 4 & 4 & 4 & 4 & 4 \\
\hline
5 & 5 & 5 & 5 & 5 & 5 \\
\hline
6 & 6 & 6 & 6 & 6 & 6 \\
\hline
7 & 6 & 6 & 6 & 6 & 7 \\
\hline
8 & 6 & 6 & 7 & 5 & 8 \\
\hline
9 & 7 & 7 & 7 & 5 & 9 \\
\hline
10 & 7 & 7 & 8 & 4 & 10 \\
\hline
11 & 8 & 8 & 8 & 4 & 11 \\
\hline
12 & 8 & 8 & 8 & 4 & 12 \\
\hline
13 & 8 & 8 & 9 & 4 & 13 \\
\hline
14 & 9 & 9 & 9 & 4 & 14 \\
\hline
15 & 9 & 9 & 9 & 4 & 15 \\
\hline
16 & 9 & 9 & 10 & 4 & 16 \\
\hline
17 & 10 & 10 & 10 & 4 & 17 \\
\hline
18 & 10 & 10 & 10 & 4 & 18 \\
\hline
\end{tabular}
\end{table}

\section{Truncation to the lowest-energy single-particle states}

\subsection{Power-law traps}

In the case of power-law traps we restricted ourselves to the lowest-Landau-level
single-particle states. This is self-consistent as long as the energy associated 
with the anharmonic part of the trapping potential and the interaction energy are 
smaller than the oscillator quantum of energy. For $m$ and $p$ of order unity, the
first energy scale is of order $\lambda$ and the second of order $g$, which have 
to be much less than unity. 

However, even if these conditions are not fulfilled, according to the scaling arguments 
of the Sec.\,V.A, one may choose a sufficiently small value of $\beta$ in such a way as 
to fulfil the two conditions.

\subsection{Hard-wall traps}

In the case of hard-wall traps, the condition that must be satisfied is that
the interaction energy remains smaller than the energy spacing between the
lowest-energy states and the first-excited states. Since $\epsilon_{n,m} = 
\alpha_{n,m}^2/({2 R^2})$, we want 
\begin{equation}
   2 \pi g \int_0^R |\psi_m^{\rm HD}|^4 \, d^2 \rho \ll \epsilon_{1,m} - \epsilon_{0,m},
\end{equation}
or
\begin{equation}
2 \pi g \int_0^R |\psi_m^{\rm HD}|^4 \, d^2 \rho \ll
\frac {1} {2 R^2} (\alpha_{1,m}^2 - \alpha_{0,m}^2).
\end{equation}
The term on the left (i.e., the interaction energy) scales also as $R^{-2}$ and 
therefore the truncation is not affected by the value of $R$.

Numerically one finds that for $m$ of order unity the interaction energy is
approximately $g/(2 R^2)$. As a result,
\begin{equation}
  g \ll (\alpha_{1,m}^2 - \alpha_{0,m}^2).
\end{equation}
From the values of $\alpha_{1,0}$ and $\alpha_{1,1}$, 
\begin{equation}
   g \ll \alpha_{1,m}^2 - \alpha_{0,m}^2 \approx 10,
\end{equation}
which safely exceeds the typical values of $g$ appearing in the phase diagram
of Fig.\,6.

\section{Structure of the phase diagram and comparison between the anharmonic and the hard-wall potentials}

We now analyse the structure of the phase diagrams. We focus on (i) the dependence of the 
instability parameter $q$ on $m_0$, (ii) the role of the trap parameters, and (iii) the 
qualitative differences between confinement types.

The method that we have adopted in Sec.\,IV allows us to examine the more subtle problem 
of the continuous transitions in great detail. Table I -- which is one of our main results 
-- summarizes the dominant instability channels (which correspond to the continuous 
transitions), highlighting the qualitative difference between power-law and hard-wall 
confinement. The values of $p$, $\lambda$, and $R$ that we used are the ones used also 
in the phase diagrams of Figs.\,1, 3, and 6. 

In the first column of this table we have the value of $m_0$, which corresponds to a 
multiply-quantized vortex state with circulation equal to ${m_0}$. As we saw, this 
state is unstable against the state of the form of Eq.\,(\ref{varphi}). The five 
columns on the right show the value of $q$ for the instability (for each $m_0$). 
Here the values of ($p$ and $\lambda$) and of $R$ (which is set equal to 3) are shown 
at the top of the table.

\subsection{Power-law traps}

First of all, in the case of an anharmonic potential, we find that for small $m_0$, 
$q = m_0$. In other words, the (leading) instability of a multiply-quantized state 
$\psi_{m_0}^{\rm LLL}$ is always against a mixed state which includes the triplet 
of states ($\psi_0^{\rm LLL}, \psi_{m_0}^{\rm LLL}$, $\psi_{2m_0}^{\rm LLL}$). 
In more physical terms, a multiply-quantized vortex state of charge $m_0$ splits 
into $2 m_0$ singly-quantized vortices, with $m_0$ of them lying closer to the 
trap center and the rest $m_0$ lying on the periphery of the cloud. In addition, 
the fact that we also have the state $\psi_0^{\rm LLL}$ in the order parameter 
of the mixed state implies that the density of the gas at the center of the trap 
does not vanish (since $\psi_0^{\rm LLL}$ is the only state which is nonzero at 
$\rho = 0$). 

On the other hand, as $m_0$ increases, $q < m_0$. The smallest value of $m_0$ 
that this is happening is $m_0 = 7$, where $q = 6$. The instability of the 
state $\psi_7^{\rm LLL}$ is against the triplet $(\psi_1^{\rm LLL}, \psi_{7}^{\rm LLL}$, 
$\psi_{13}^{\rm LLL}$). In this case, we therefore have thirteen vortices, with 
one being at the center of the trap, six (singly-quantized) closer to the center 
and the rest six (singly-quantized) residing further away from the center. 

In Fig.\,1 we set $p=2$ and examine the effect of the strength of the anharmonic 
term $\lambda$, on the phase diagram. Here $\lambda = 0.05$ in the higher plot and
$\lambda = 0.08$ in the lower one. As expected (according to the results of Sec.\,III), 
the slope of the solid (red) lines, which indicate the discontinuous 
transitions, is negative. Also, since $p=2$, the spacing (for $g=0$) between successive lines 
is equal to $\lambda$. Finally, the curves which give the continuous phase transitions, 
dotted (black) curves shift upwards as $m_0$ increases. 

As $m_0$ increases further, as the results of Table I show, for ($p=2$ and $\lambda 
= 0.05$) and ($p=2$ and $\lambda = 0.08$) and also for ($p=3$ and $\lambda = 0.001$), 
$q$ continues to increase. 

For ($p = 2$ and $\lambda = 0.05$), and for ($p=2$ and $\lambda = 0.08$), then 
$q = 9$. In these cases, the multiply-quantized vortex state $\psi_{16}^{\rm LLL}$ 
splits to the triplet of states $(\psi_7^{\rm LLL}, \psi_{16}^{\rm LLL}, 
\psi_{25}^{\rm LLL})$. Here we have a multiply-quantized vortex state with multiplicity 
equal to seven at the center of the trap. Another nine singly-quantized vortices are 
closer to the trap center, while the rest nine are further away. Figure 2 shows the 
density and the phase of the order parameter for $\psi_{16}^{\rm LLL}$ and for the
triplet $(\psi_7^{\rm LLL}, \psi_{16}^{\rm LLL},\psi_{25}^{\rm LLL})$. 

In Fig.\,3 we fix $\lambda$ to the value 0.001 and consider two different values 
of $p = 3$ (higher) and $p=4$ (lower). Again, the slope of the solid (red)
lines is negative, and becomes more negative with increasing $m_0$. Also, the 
spacing (for $g=0$) between successive lines increases, as both $m_0$ 
and $p$ increase, in agreement with Eq.\,(\ref{rvsom11}). 

For ($p = 3$ and $\lambda = 0.001$), $q = 10$ and therefore the state 
$\psi_{16}$ is unstable against $(\psi_6^{\rm LLL}, \psi_{16}^{\rm LLL}, 
\psi_{26}^{\rm LLL})$. In the mixed state we have a multiply-quantized vortex 
state with a multiplicity equal to six at the center of the trap. Another ten 
singly-quantized vortices are closer to the trap center, while the rest ten 
(singly-quantized) vortices are further away. Figure 4 shows the density and 
the phase of the corresponding order parameter. 

When ($p = 4$ and $\lambda = 0.001$) Table I shows that $q = 4$. In this case, 
the multiply-quantized vortex state $\psi_{16}^{\rm LLL}$ is unstable against 
the triplet of states $(\psi_{12}^{\rm LLL}, \psi_{16}^{\rm LLL}, \psi_{20}^{\rm LLL})$. 
Therefore, we have a multiply-quantized vortex state at the trap center with 
multiplicity equal to twelve, four singly-quantized vortices closer to the trap 
center and four more singly-quantized vortices further away from the center of 
the trap, as Fig.\,5 illustrates. More generally, for ($p = 4$ and $\lambda = 
0.001$), according to Table I, $q$ has a non-monotonic behaviour as a function 
of $m_0$. Initially it increases, reaches a maximum, and then decreases 
and saturates to some constant value, $q = 4$. 

\subsection{Hard-wall traps}

Turning to the hard-wall potential, as Table I shows, here $q = m_0$, for all the 
values of $m_0$ that we considered (contrary to the case of an anharmonic 
potential). In Fig.\,6 we show the phase diagram for a hard-wall potential, with 
$R = 2$, 3, and 4, from top to bottom. Here, for large values of $m_0$ the slope 
of the solid (red) lines, which correspond to the discontinuous 
transitions is positive, see Eq.\,(\ref{rvsom12}), contrary to the anharmonic potentials. 
Also, in agreement with Eq.\,(\ref{rvsom122}), the spacing between successive  
lines tends to a constant, for large values of $m_0$. As mentioned also in Sec.\,V, 
the top and the bottom plots, for $R=2$ and $R=4$, are identical (up to a rescaling
of the horizontal axis) due to the scaling argument that we described in Sec.\,V. 

Figure 7 shows the density and the phase of the order parameter, where the 
multiply-quantized vortex state $\psi_{16}^{\rm HD}$ is unstable against the triplet of 
states $(\psi_{0}^{\rm HD}, \psi_{16}^{\rm HD}, \psi_{32}^{\rm HD})$. 

\begin{figure}[t]
\includegraphics[width=7cm,height=6cm,angle=0]{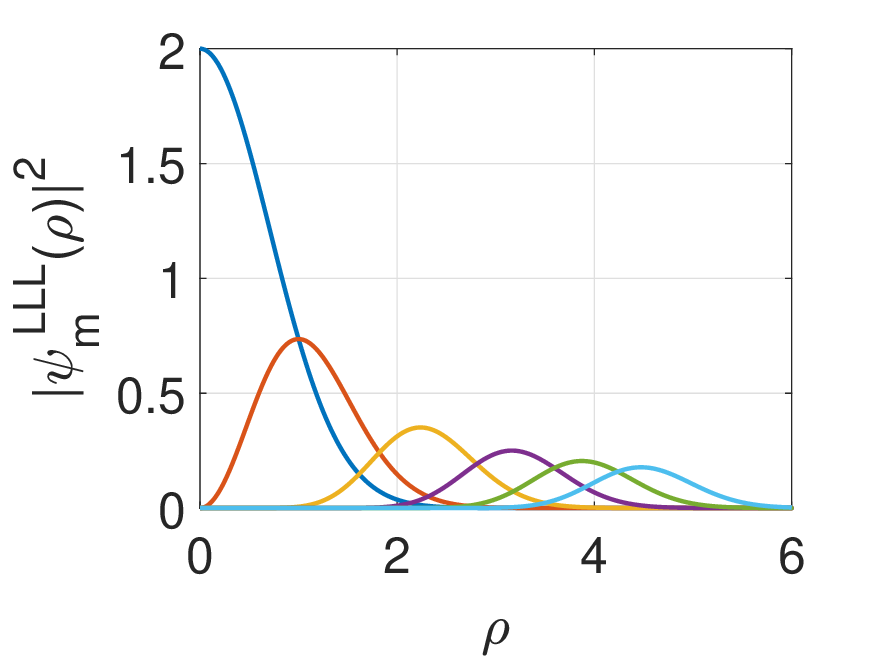}
\includegraphics[width=7cm,height=6cm,angle=0]{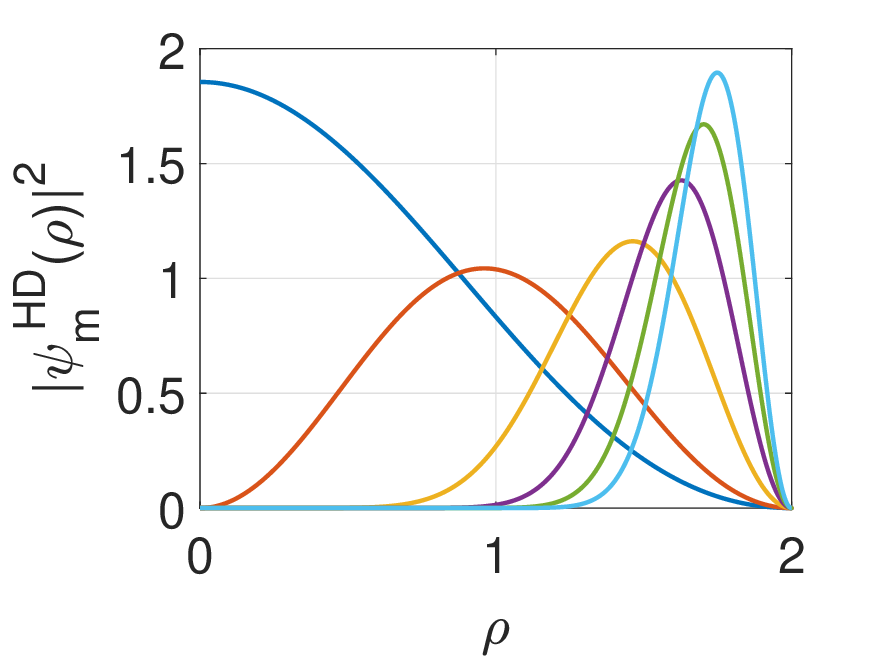}
\caption{The radial density that corresponds to the eigenstates of the 
lowest-Landau states $|\psi_{m}^{\rm LLL}(\rho)|^2$ (higher) and of the Bessel eigenstates, 
$|\psi_{m}^{\rm HD}(\rho)|^2$, with $m = 0, 1, 5, 10, 15$, and 20, from left to right. The 
$x$ axis is the radial coordinate $\rho$, measured in units of $a_0$ in the higher plot 
and in units of $R_0$ in the lower one. The $y$ axis is measured in units of $a_0^{-2}$ 
in the higher plot and in units of $R_0^{-2}$ in the lower plot, where $R=2$.}
\end{figure}

\subsection{Comparison between the two types of confinement}

From the results of Table I we see that for the anharmonic potential and for $m_0 \le 6$, 
then $q = m_0$, for all the values of $\lambda$ and $p$ that we have considered. For 
$m_0 > 6$, depending on the parameters we have chosen, $q$ either continues to increase 
(first three columns), or it decreases, then becoming equal to $q=4$ (fourth column). 

When $q = m_0$, the mixed state includes $\psi_0^{\rm LLL}$ and the density of the cloud
at the trap center is nonzero. For sufficiently large values of $m_0$, $\psi_0^{\rm LLL}$ 
is no longer in the triplet of states of the mixed state and as a result the cloud develops 
a hole at the trap center. This is due to the effective potential, 
\begin{equation}
  V_{\rm eff}(\rho) = V_{\perp}(\rho) - \frac 1 2 \Omega^2 \rho^2
  = \frac 1 2 (1 - \Omega^2) \rho^2 + \frac 1 2 \lambda {\rho}^{2p},
\end{equation}
which has a Mexican-hat shape for $\Omega>1$. The minimum of $V_{\rm eff}(\rho)$ is 
at a distance 
\begin{equation}
  \rho_0 = \left[ \frac 1 {\lambda p} \left( \Omega^2 - 1\right) 
  \right]^{1/(2p-2)}
  \label{rho0}
\end{equation} 
and is an increasing function of $\Omega$, and, correspondingly, of $m_0$. For
$\Omega \approx 1$, 
\begin{equation}
  \rho_0 \approx \left[ \frac 2 {\lambda p} \left( \Omega - 1 \right) \right]^{1/(2p-2)}.
\end{equation}
From Eq.\,(\ref{neqqq}) it follows that 
\begin{equation}
   m_0 \approx \left[ \frac 2 {\lambda p} (\Omega - 1) \right]^{1/(p-1)}.
\end{equation}
Therefore $\rho_0 \approx \sqrt{m_0}$, which is consistent with the maximum of 
$|\psi_{m_0}^{\rm LLL}(\rho)|^2$ that occurs at $\rho = \sqrt{m}$. In other words, 
the density of the cloud adjusts to the minimum of $V_{\rm eff}(\rho)$ and, as a 
result, it spreads out more and more, while at the same time the average density 
drops. The upper plot in Fig.\,8 shows $|\psi_m^{\rm LLL}(\rho)|^2$ that corresponds 
to the lowest-Landau level eigenstates, with $m = 0, 1, 5, 10, 15$, and 20.

For a hard-wall potential, $q = m_0$ for all the values that we have considered, 
contrary to the anharmonic potential. In this case, the effective potential has the 
form of an inverse parabola,
\begin{equation}
  V_{\rm eff}(\rho) = - \frac 1 2 \Omega^2 \rho^2,
\end{equation}
with $0 \le \rho < R$. Obviously its minimum occurs at $\rho_0 = R$, i.e., the cloud 
is pushed towards the edge of the hard-wall potential. According to Eq.\,(\ref{qqee}),
$m_0 \approx \Omega R^2$ in this case. This contrasts with the power-law case, where 
$m_0$ is set by the position of the minimum of the effective potential, due to the 
form of the power-law trap. The lower plot in Fig.\,8 shows $|\psi_m^{\rm HD}(\rho)|^2$ 
that corresponds to the Bessel eigenstates, with $n=0$, and $m = 0, 1, 5, 10, 15$, and 20.

Given the above results, we can get a better understanding of the instability as 
arising from the competition between the curvature (discrete second derivative) 
of the single-particle spectrum, $\epsilon_{m_0-q} - 2 \epsilon_{m_0} + \epsilon_{m_0+q}$ 
with respect to $m_0$, and the interaction energy. The single-particle contribution 
favours the occupation of a multiply-quantized vortex state, while the interaction 
energy favours spatial redistribution of the density through mixed states.
 
Starting with the single-particle contribution, in the anharmonic potential and for 
large $m_0$ and $p$ of order unity, the curvature scales as $\lambda p (p-1) m_0^{p-2}$, 
i.e., it depends strongly on $m_0$. In contrast, for the hard-wall potential it is equal 
to $q^2/R^2$, and is independent of $m_0$. 

The interaction energy (being repulsive) favours a homogeneous density distribution.
In the anharmonic trap, setting $q=m_0$, i.e., choosing the highest allowed value of 
$q$, leads to a large single-particle energy cost and does not reduce the interaction 
energy. Indeed, the density of $|\psi_{m}^{\rm LLL}(\rho)|^2$ is peaked at $\rho = 
\sqrt{m}$ (see, also, the higher plot in Fig.\,8), so the components $(\psi_0^{\rm LLL}, 
\psi_{m_0}^{\rm LLL}, \psi_{2m_0}^{\rm LLL})$ are localized at different radii. This 
results in a strongly inhomogeneous density distribution (radially), making such a state 
energetically unfavourable. As a result, the system finds its lowest-energy state 
choosing some intermediate value, $q < m_0$.

In contrast, in the hard-wall potential, when the system undergoes the transition to 
a mixed state, it chooses the value of $q = m_0$, which involves the triplet of states 
$(\psi_0^{\rm HD}, \psi_{m_0}^{\rm HD}, \psi_{2m_0}^{\rm HD})$. For large $m_0$, the 
eigenstates are localized near the boundary $\rho=R$, whereas the state with $m=0$ is 
spread throughout the system (see the lower plot in Fig.\,8). As a result, the state 
$(\psi_0^{\rm HD}$ is energetically favoured by interactions, since it produces a more 
uniform radial density. In this way, it pays a single-particle energy cost, i.e., 
$m_0^2/R^2$, but gains interaction energy.

\section{Experimental relevance} 

The theoretical phases identified in this study are within reach of current cold-atom 
experiments. The tuning of the coupling constant $g$ via Feshbach resonances allows for 
the exploration of the vertical axis of our phase diagrams. Furthermore, the transition 
from harmonic to anharmonic \cite{Bretin2004, Perrin, Perrin2}, or hard-wall confinement 
\cite{Gaunt2013} has already been demonstrated experimentally.

The predicted splitting of a multiply-quantized vortex into a ``necklace" of 
singly-quantized vortices could be observed using standard absorption imaging, or 
time-of-flight expansion. To probe the circulation and verify the structure 
of the predicted vortex states, one can utilize standard time-of-flight expansion imaging. 
This technique has been analysed experimentally and theoretically in Ref.\,\cite{Murray}, 
who demonstrated that releasing a ring-shaped condensate from its trap causes it to expand 
rapidly inward, where the minimum radius of the filled-in central hole serves as a direct, 
robust diagnostic of the initial quantization. The stability of these states against the 
predicted instability could provide a clear signature for identifying the underlying 
power-law index $p$ of the trapping potential through rotational response measurements. 
The presence or absence of a central density hole at the trap center is a very well-defined 
prediction, which provides a direct experimental signature. 

Let us now make some estimates regarding the actual parameters in a real experiment.
In an anharmonic potential \cite{Bretin2004, Perrin, Perrin2}, for a typical atom number 
$N = 10^5$ and a width of the condensate on the $xy$ plane 10 $\mu$m, the width $Z$ should 
be smaller, in order for the system to be quasi-two-dimensional. Considering, for example, 
$ Z = 1$ $\mu$m, the typical density per unit length $\sigma \sim 10^9$ cm$^{-1}$. Then, 
for a typical value of the scattering length $a \sim 100$ \AA, $\sigma a \sim 10^3$. 
Since we want $\sigma a$ to be smaller than unity, one should reduce the atom number, 
say $N \sim 10^2$. Last, but not least, the energy that is associated with the anharmonic
term in the trapping potential has to be smaller than both the energy of the harmonic 
potential and the interaction energy. For $\sigma a \sim 1$, $\lambda$ should be less
than $2 m^{1-p}$, for large $m$ and for $p$ of order unity. For, e.g., $m = 20$ and 
$p=2$ (i.e., in a quartic potential), $2 m^{1-p} = 10^{-1}$, while in the experiments 
of Refs.\,\cite{Bretin2004, Perrin}, $\lambda \approx 10^{-3}$.  

In a hard-wall potential \cite{Gaunt2013}, for a typical number of atoms $N \sim 10^5$ 
and $R_0 \sim 10$ $\mu$m, in order for the system to be in the quasi-two-dimensional 
limit, the width of the condensate along the axis of rotation, $Z$, would have to be 
smaller than $R_0$. If, for example, $Z \sim 1$ $\mu$m, then the density per unit length 
$\sigma$ is roughly $10^{9}$ cm$^{-1}$, as in the previous case. For a typical value of 
the scattering length $a \sim 100$ \AA, then $\sigma a \sim 10^3$. This implies that one 
should use a substantially smaller number of atoms, $N$, on the order of $10^3$. In this 
case, $\sigma a$ would be on the order of 10. Regarding the value of $\Omega$, this is 
set by the $\hbar/(M R_0^2)$. For a typical mass atom mass $M$ of 100 proton masses,  
the characteristic scale of the rotational frequency $\Omega$ is a few Hz.  

\section{Summary}

In this work, we derived the rotational phase diagram of a Bose-Einstein condensate confined 
in power-law and hard-wall potentials, identifying the mechanisms that govern the stability 
of vortex states. 

For weak interactions, the system exhibits discontinuous transitions between multiply-quantized 
vortex states as the rotation frequency increases. As the interaction strength grows, these states 
become unstable through continuous transitions toward mixed configurations involving both singly 
and multiply-quantized vortices. The onset of these instabilities is controlled by the competition 
between the curvature of the single-particle spectrum and the interaction energy, which favours 
spatial redistribution of the density.

A key result of this study is the qualitative difference between power-law and hard-wall 
confinement. In hard-wall potentials, the leading instability always involves the single-particle
state with zero angular momentum, resulting in a nonzero density at the trap center. In contrast, 
in power-law traps, the effective potential develops a Mexican-hat profile at sufficiently large 
rotation frequencies, leading to a depletion of the density at the trap center.

We have also demonstrated that the derived phase diagram for both classes of trapping potentials 
exhibits universal scaling properties. In power-law traps, a rescaling of the parameters 
$1 - \Omega$, $\lambda$, and $g$ leaves the phase diagram invariant, while in hard-wall traps 
the phase diagram depends only on the combination $\Omega R^2$. 

An anharmonic trapping potential with a large power-law exponent $p$ approaches a hard-wall 
potential. However, the physical behaviour we obtain differs from that of a true hard-wall 
trap. This difference arises because our analysis assumes a weak anharmonic term and uses 
the lowest-Landau-level states. This, however, does not capture the physics of a true hard-wall 
potential, even for large $p$.

Our results are directly relevant to current cold-atom experiments, where both the interaction 
strength and the trapping geometry can be tuned with high precision. The predicted vortex-splitting 
mechanisms, the presence or absence of a central density hole, and the structure of the phase 
boundaries provide well-defined theoretical predictions, which are worth investigating 
experimentally.

\begin{acknowledgements}

The author wishes to thank Andrew Jackson for useful discussions.

\end{acknowledgements}

\end{document}